\documentclass[a4paper, 12pt, oneside]{article}
\usepackage[cp1251]{inputenc} 
\usepackage[english]{babel} 
\usepackage{ graphics,amsfonts, amsmath,bm,amssymb, amsmath}
\usepackage[dvips]{graphicx}
\usepackage[flushleft,small]{caption2} 
\renewcommand{\@biblabel}[1]{#1.\hfill}
\newcommand{\diag}{\rm \diag\, }

\renewcommand{\Re}{\mathop{\rm Re\,}}
\renewcommand{\Im}{\mathop{\rm Im\,}}

\makeatother {

\begin{document}
 \thispagestyle{empty}
\large
\renewcommand{\refname}{\begin{center}\rm
REFERENCES
\end{center}}

\begin{center}
STRUCTURE OF THE ELECTRIC FIELD IN THE SKIN EFFECT PROBLEM
\end{center}
\begin{center}
 {\bf Yu.F. Alabina}
\end{center}

\begin{center}
 Moscow State Regional University\\
{\it 105005, Moscow, Radio st., 10 a\\
e-mail: yf.alabina@gmail.com}
\end{center}

\begin{abstract}
The structure of the electric field in a plasma has been elucidated
for the skin effect problem. An expression for the distribution
function  in the half-space and the electric field profile have been
obtained in the explicit form. The absolute value, the real part,
and the imaginary part of the electric filed have been analyzed in
the case of the anomalous skin effect near to a plasma resonance. It
has been demonstrated that the electric field in the skin effect
problem is predominantly determined by the discrete spectrum, i.e.,
the oscillation frequency of external field is the value of plasma
frequency.

\end{abstract}

{\bf 1. Introduction. Statement of problem.}

The skin effect is associated with the response of an electron gas
(in a metal or in a gas plasma) to an external alternating
electromagnetic field that is tangential to the surface \cite{1,2}.
This classical problem has been studied by many authors \cite{3} --
\cite{6} and, up to now, has remained the subject of investigation.
The main attention has been focused on the calculation of the
impedance. The distribution function of electrons and the electric
field in plasma almost have not been investigated previously.

It has been demonstrated that the electric field is the sum of the
integral term and two (or one) exponentially decreasing particular
solutions to the initial system and that one particular solution
disappears depending on the anomaly parameter.

Let's Maxwell plasma fills the half-space $x>0$. Here $x$ is the
orthogonal coordinate  to the plasma boundary. Let's the external
electric field has only  $y$ component. Then the self-consistent
electric field inside in plasma also has only $y$ component
$E_y(x,t)=E(x)e^{-i\omega t}$. We consider the kinetic equation for
the electron distribution function:
$$
\dfrac{\partial f}{\partial t}+\text{v} _{x}\dfrac{\partial
f}{\partial x}+eE(x)e^{-i\omega t}\dfrac{\partial f}{\partial
p_y}=\nu(f_0-f(t,x,\mathbf{v})). \eqno{(1)}
$$

In (1) $\nu$ is the frequency of electron collisions with ions, $e$
is the charge of electron, $f_0(\text{v})$ is the equilibrium
Maxwell distribution function, $\textbf{p}=m\text{v}$ is the
momentum of electron,
$$
f_0(\text{v})=n\left(\dfrac{\beta}{\pi}\right)^{3/2}
\exp(-\beta^2\text{v}^2),\quad \beta=\dfrac{m}{2k_BT}.
$$

Here  $m$ is the mass of electron, $k_B$ is the Boltzmann constant,
$T$ is the temperature of plasma, $\text{v}$ is the modulus of the
velocity of the electron, $n$ is the concentration of electrons
(number density), $c$ is the speed of light.

The electric field $E(x)$  satisfies to the equation:
$$
E''(x)= -\dfrac{4\pi i  e^{i\omega t}\omega e}{c^2} \int v_y
f(t,x,\mathbf{v})\,d^3v. \eqno{(2)}
$$

We assume that intensity of an electric field is such that linear
approximation is valid. Then distribution function can be presented
in the form:
$$
f=f_0\left(1+C_y\exp(-i\omega t)h(x,\mu)\right),
$$
where $\textbf{C}=\sqrt{\beta}\text{v}$ is the dimensionless
velocity of electron, $\mu=C_x$. Let $l=v_T\tau$ is the mean free
path of electrons, $v_T=1/\sqrt{\beta}$, $v_T$ is the thermal
electron velocity,\; $\tau=1/\nu$. We introduce the dimensionless
parameters and the electric field:
$$
t_1=\nu t,\quad x_1=\dfrac{x}{l}, \quad
e(x_1)=\dfrac{\sqrt{2}e}{\nu\sqrt{mk_BT}}E(x_1).
$$

Later we substitute $x_1$ for $x$. The substitution produces the
following form of the kinetic equation (1) and the equation on a
field with the displacement current (2):
$$
\mu\dfrac{\partial h}{\partial x}+z_0\,h(x,\mu)=e(x)
 \eqno{(3)}
$$
$$
e''(x)=-i\dfrac{\alpha}{\sqrt{\pi}}
\int\limits_{-\infty}^{\infty}\exp(-{\mu'}^2)\,h(x,\mu')\,d\mu'.
\eqno{(4)}
$$

Here,
$$\alpha=\dfrac{2l^2}{\delta^2},  \quad z_0=1-i\Omega,
 \quad \Omega=\omega\tau=\dfrac{\omega}{\nu}, \quad \delta=\dfrac{c^2}{2\pi\omega\sigma_0},$$
$\delta$ is the classical depth of the skin layer \cite{1},
$\sigma_0=\dfrac{e^2n}{m\nu}$, $\sigma_0$ is the electric
conductance,  $\alpha$ is the anomaly parameter.

The boundary conditions at the plasma surface for the distribution
function of electrons in the case of specular reflection of
electrons from the boundary can be written as follows \cite{1}:
$$
h(0,\mu)=h(0,-\mu), \qquad 0<\mu<\infty. \eqno{(5)}
$$

The distribution function will be sought in the form of a decaying
function far from the boundary; that is,
$$
h(+\infty,\mu)=0, \qquad -\infty<\mu<0. \eqno{(6)}
$$

The electric field deep in the plasma far from the surface decays.
Taking into account this circumstance, the boundary conditions for
the electric field are written in the form
$$
e(0)=1, \eqno{(7)}
$$
$$
e(+\infty)=0. \eqno{(8)}
$$

{\bf 2. Decomposition on eigenfunctions}

The separation of variables in (3) and (4) within several steps
leads to the exponentially decreasing solutions
$$
h_\eta(x,\mu)=\exp(-\dfrac{z_0x}{\eta})\Phi(\eta,\mu), \quad
e_\eta(x)=\exp(-\dfrac{z_0x}{\eta})E(\eta), \eqno{(9)}
$$
where the separation parameter (also termed the spectral parameter)
$\eta$ continuously fills the interval $(0, \infty)$, which,
therefore, is called the continuous spectrum of the problem.

Substitution of relationships (9) into the initial system of
equations (3) and (4) leads to the characteristic system of
equations
$$
(\eta-\mu)\Phi(\eta,\mu)=\dfrac{\eta}{z_0} E(\eta),
$$
$$
\dfrac{z_0^2}{\eta^2}E(\eta)=-i\dfrac{\alpha}{\sqrt{\pi}}
\int\limits_{-\infty}^{\infty} \exp(-\mu^2)\Phi(\eta,\mu)\,d\mu.
$$

The functions $\Phi(\eta,\mu)$ and $E(\eta)$, which are referred to
as the eigenfunctions of the characteristic system and correspond to
the eigenvalue (or characteristic value) of the parameter $\eta$,
are defined by the expressions
$$
\Phi(\eta,\mu)=\dfrac{a}{\sqrt{\pi}}\eta^3P\dfrac{1}{\eta-\mu}+
\lambda(\eta)\exp(\eta^2)\delta(\eta-\mu),
\eqno{(10a)}
$$
$$
E(\eta)=\dfrac{az_0}{\sqrt{\pi}}\eta^2,
\eqno{(10b)}
$$
where the dispersion function $\lambda(z)$ (see, for example,
\cite{3}) is given by the formula
$$
\lambda(z)=1+\dfrac{a z^3}{\sqrt{\pi}}
\int\limits_{-\infty}^{+\infty}\dfrac{\exp(-\mu^2)}{\mu-z}\,d\mu,
\qquad a=-\dfrac{i\alpha}{z_0^3}. \eqno{(11)}
$$

With the use of the argument principle, it is possible to show that,
in the $(\alpha, \Omega)$ plane, there exists a domain $D^+$ (Fig.
1a) so that, if the point $(\alpha, \Omega) \in D^+$, the dispersion
function has four zeros $\pm \eta_0$ and $\pm \eta_1$, and if
$(\alpha, \Omega) \in D^-$ (where $D^-$ is exterior of the domain
$D^+$), the dispersion function has two zeros $\pm \eta_0$. The
designations $\eta_0$ and $\eta_1$ correspond to the zeros with the
positive real parts: $\Re \eta_0>0$ and $\Re \eta_1 > 0$. The
boundary of the domain $D^+$ is found from the equation $\omega^\pm
(\mu) = 0$ and, in the parametric form, is determined by the
equations $1-3\Omega^2\pm \alpha\, q(\mu)=0,\;
3\Omega-\Omega^3-\alpha \,p(\mu)=0,\; -\infty<\mu<+\infty$.

 It
should be noted that the parameters $\alpha$ and $\Omega$ are
proportional to the electric field frequency; i.e., they are not
independent. In this respect, it seems quite natural to introduce
the dimensionless independent frequencies
$$
\omega_1=\dfrac{\omega}{\omega_p v_c},\qquad \quad
\nu_1=\dfrac{\nu}{\omega_p\,v_c},
$$
and to construct the corresponding domains $D^+_1$ and $D^-_1$ (Fig.
1b) in their plane. Here, $\alpha=\omega_1/(\nu_1^3), \;
\Omega=\omega_1/\nu_1$,\quad $\omega_p$ is the plasma frequency  and $n$
is the electron concentration.

The zeros $\eta_0$ and $\eta_1$ correspond to the following
eigenfunctions of the characteristic equation that are associated
with the discrete spectrum:
$$
\Phi(\eta_k,\mu)=\dfrac{a\eta_k^3}{\sqrt{\pi}(\eta_k- \mu)}, \qquad
E(\eta_k)=\dfrac{a z_0 \eta_k^2}{\sqrt{\pi}}, \qquad k=0,1.
$$

The zeros of the dispersion function can be calculated in the
explicit form with the use of the formulas for its factorization. In
the case of two zeros $\pm \eta_0$, the dispersion function (see
\cite {3}) can be represented in the form
$$
\lambda(z)=a(\eta_0^2-z^2)X(z)X(-z),
$$
where
$$
X(z)=\exp V(z),
$$
$$
V(z)=\dfrac{1}{2\pi i}
\int\limits_{0}^{\infty}\dfrac{\ln G(\tau)\,d\tau}{\tau-z},
$$

In the case of four zeros $\pm \eta_0$ and $\pm \eta_1$, the
dispersion function can be written as follows:
$$
\lambda(z)=a(\eta_0^2-z^2)(\eta_1^2-z^2)X_1(z)X_1(-z),
$$
where the function $X_1(z)$ is expressed through the function
$X(z)$: $X_1(z)=X(z)/(z-1)$.

By calculating the left- and right-hand sides of the former formula
for the factorization of the dispersion function (for, example, at
the point $z = 0$), after some transformations, we obtain the
relationship for its zeros
$$
\pm \eta_0=\dfrac{1}{\sqrt{aX(z)X(-z)}}+1.
$$

In the skin effect theory, the normal and anomalous skin effects are
recognized \cite{7}. In the case of the normal skin effect, the mean
free path of electrons is considerably smaller than the skin depth;
i.e., the anomaly parameter satisfies the inequality $\alpha\ll 1$.
The anomalous skin effect corresponds to the case where the mean
free path of electrons is considerably larger than the
characteristic skin depth: $\alpha\gg 1$.

Let us construct the general solution to the initial system of
equations in the form of the expansion in eigenfunctions of the
discrete and continuous spectra. Since the discrete spectra for zero
and unit indices are different and the continuous spectrum does not
depend on the index, the expansions of the solution in both cases
differ only in the nonintegral terms corresponding to the discrete
spectrum.

In \cite{5}, it was demonstrated that the distribution function of
electrons and the electric field, which are the solution to the
problem described by expressions (3)--(8), have the following
expansions:
$$
h(x,\mu)=\dfrac{a}{\sqrt{\pi}}\sum\limits_{k=0}^1
\dfrac{A_k\,\eta_k^3}
{\eta_k-\mu}\exp\left( -\dfrac{z_0\,x}{\eta_k}\right)+
$$
$$
\quad  \quad \quad \quad\quad+
\int\limits_{0}^{\infty} \exp\left( -\dfrac{z_0\,x}{\eta}\right)
A(\eta)\Phi(\eta,\mu)\,d\eta,
\eqno{(12)}
$$

$$
e(x)=\dfrac{az_0}{\sqrt{\pi}}\sum\limits_{k=0}^1A_k\eta_k^2
\exp\Big(-\dfrac{z_0x}{\eta_k}\Big)+
$$
$$\qquad \qquad+
\dfrac{az_0}{\sqrt{\pi}}\int\limits_{0}^{\infty}
\exp\Big(-\dfrac{z_0x}{\eta}\Big)\eta^2A(\eta)\,d\eta.
\eqno{(13)}
$$

Here, $\Re \eta_k>0,\; A_k\; \;(k=0,1)$ are unknown constant
coefficients of expansions (12) and (13) (the so-called coefficients
of the discrete spectrum), and $A(\eta)$ is an unknown function (the
so-called coefficient of the continuous spectrum).

It should be noted that, in the case of two zeros of the dispersion
function, it is necessary to set $k = 0$ in relationships (12) and
(13). Therefore, the structure of the electric field depends on the
domain $D^\pm$ that contains the point with the parameters $(\alpha,
\Omega)$.

In \cite{8}, it was shown that the coefficient $A(\eta)$ of the
continuous spectrum is represented in the form
$$
A(\eta)= -\dfrac{\eta \exp(-\eta^2)}
{z_0\,I\,\lambda^+(\eta)\lambda^-(\eta)},
$$
{where}
$$\quad
I=\dfrac{1}{2\pi}\int\limits_{-\infty}^{\infty}\dfrac{d\tau}
{\lambda(i\tau)}.
$$

The coefficients of the discrete spectrum are written in the
following form:
$$
A_k= -\dfrac{\sqrt{\pi}}{az_0I\eta_k^2 \lambda'(\eta_k)}, \qquad
k=0,1.
$$

The impedance is given by the formula \cite{1}
$$
Z=\dfrac{4\pi i \omega}{c^2}\cdot\dfrac{e(0)}{e'(0)}.
$$

According to the boundary conditions for the filed, we have
$e(0)=1$. Therefore, the following expression holds true for the
impedance:
$$
Z=\dfrac{4\pi i \omega}{c^2e'(0)}= \dfrac{8\pi i \omega
l}{c^2z_0}\Bigg[\dfrac{1}{\pi} \int\limits_{0}^{\infty}
\dfrac{d\tau}{\lambda(i\tau)}\Bigg].
$$

{\bf 3. Distribution function and the electric field}

With the use of the determined coefficients of the continuous and
discrete spectra, the electric field profile in the half-space can
be represented in the explicit form
$$
e(x)=-\dfrac{1}{I\lambda'(\eta_0)}
\exp\Big(-\dfrac{z_0x}{\eta_0}\Big)-
\dfrac{1}{I\lambda'(\eta_1)}\exp\Big(-\dfrac{z_0x}{\eta_1}\Big)-
$$$$
-\dfrac{a}{I\sqrt{\pi}}\int\limits_{0}^{\infty}
\exp\Big(-\dfrac{z_0x}{\eta}\Big)\dfrac{\eta^3\exp(-\eta^2)}{
\lambda^+(\eta)\lambda^-(\eta)}\,d\eta. \eqno{(14)}
$$

Formula (14) will be subsequently used for analyzing the behavior of
the electric field in the half-space.

Now, we consider the profile of the distribution function of
electrons in the half-space in the explicit form. The distribution
function is represented in the form of two terms:
$$
h(x,\mu)=h_d(x,\mu)+h_c(x,\mu),
$$
where the terms $h_d(x,\mu)$ and $h_c(x,\mu)$ correspond
to the discrete and continuous
spectra, respectively. With the use of the equality for the
coefficients of the discrete and continuous spectra, these terms are
written as follows:
$$
h_d(x,\mu)=-\dfrac{1}{z_0I}
\sum\limits_{k=0}^{1}\dfrac{\eta_k}{(\eta_k-\mu)\lambda'(\eta_k)}
\exp\Big(-\dfrac{z_0x}{\eta_k}\Big),
$$
$$
h_c(x,\mu)=-\dfrac{1}{z_0I}\int\limits_{0}^{\infty}\exp
\Big(-\dfrac{z_0x}{\eta}\Big)\dfrac{\eta\exp(-\eta^2)}{\lambda^+(\eta)
\lambda^-(\eta)}\Phi(\eta,\mu)\,d\eta.
$$

At the plasma boundary, i.e., at $x = 0$, the last relationship can
be calculated in the explicit form. As a result, we have
$$
h_c(0,\mu)=\dfrac{1}{2\pi iz_0I}\int\limits_{0}^{\infty}
\Big(\dfrac{1}{\lambda^+(\eta)}-\dfrac{1}{\lambda^-(\eta)}\Big)
\dfrac{\eta\,d\eta}{\eta-\mu}+
$$
$$
+\dfrac{\lambda(\mu)\theta(\mu)\exp(\mu^2)}{2iz_0Ia\sqrt{\pi} \mu^2}
\Big(\dfrac{1}{\lambda^+(\mu)}-\dfrac{1}{\lambda^-(\mu)}\Big),\quad
\mu\in (-\infty,+\infty).
$$
Here, $\theta(\mu)=1$ at $0<\mu<\infty$ and $\theta(\mu)=0$ at
$-\infty<\mu<0$.

By using contour integration methods \cite{5}, for the first term we
find that
$$
h_c(0,\mu)=\dfrac{1}{z_0I}\Bigg[\dfrac{\mu\lambda(\mu)(1-
\theta(\mu))}{\lambda^+(\mu)\lambda^-(\mu)}+
\dfrac{\eta_0}{(\eta_0-\mu)\lambda'(\eta_0)}+
$$
$$
+\dfrac{\eta_1}{(\eta_1-\mu)\lambda'(\eta_1)}+
\dfrac{1}{\pi}\int\limits_{0}^{\infty}\dfrac{\tau^2\,d\tau}
{\lambda(i\tau)(\tau^2+\mu^2)}\Bigg].
$$

By summing up the terms corresponding to the discrete and continuous
spectra in the distribution function, we finally obtain
$$
h(0,\mu)=\dfrac{1}{z_0I}\Bigg[\dfrac{\mu\lambda(\mu)
(1-\theta(\mu))}{\lambda^+(\mu)\lambda^-(\mu)}+
\dfrac{1}{\pi}\int\limits_{0}^{\infty}\dfrac{\tau^2\,d\tau}
{\lambda(i\tau)(\tau^2+\mu^2)}\Bigg].
$$

From this expression, for the distribution function of electrons
moving at the metal boundary (i.e., in the case $-\infty<\mu<0$), we
have
$$
h(0,\mu)=\dfrac{1}{z_0I}\Bigg[\dfrac{\mu\lambda(\mu)
}{\lambda^+(\mu)\lambda^-(\mu)}+
\dfrac{1}{\pi}\int\limits_{0}^{\infty}\dfrac{\tau^2\,d\tau}
{\lambda(i\tau)(\tau^2+\mu^2)}\Bigg], \quad -1<\mu<0,
$$

For electrons specularly reflected from the metal boundary, we
derive
$$
h(0,\mu)=\dfrac{1}{\pi\, z_0\,I}
\int\limits_{0}^{\infty}\dfrac{\tau^2\,d\tau}
{\lambda(i\tau)(\tau^2+\mu^2)}, \quad 0<\mu<1.
$$

These functions satisfy the specular boundary condition
$h(0,\mu)=h(0,-\mu)$.

For all subsequent figures, we consider the typical case with the
ratio $v_F/c=0.003$.

The behavior of the real and imaginary parts of the distribution
function at the boundary is illustrated in Fig. 2. In view of the
specular boundary condition, the distribution functions of electrons
reflected from the boundary ($0<\mu<\infty$) and electrons moving to
the boundary ($-\infty<\mu<0$) are symmetric with respect to the
point $\mu=0$. The functions are constructed for the parameters
\;$\alpha=1, \Omega=333$. Figure 2a depicts the real part of the
electric field,  fig. 2b depicts the imaginary part of the electric
field. Let us compare the imaginary part of the electric field $\Im
e_c(x)$ with the real part of the field $\Re e_c(x)$. The
electric-field amplitude $|\Im(h(0,\mu))|>6\cdot 10^3$ is
considerably bigger for the imaginary part. It should be noted that
the distribution function rapidly decreases with an increase in the
quantity $\mu$. This circumstance is a manifestation of the
ineffectiveness concept \cite{2}, according to which only electrons
moving almost parallel to the surface, i.e., for which the quantity
$\mu$ is considerably smaller than unity, are significant in the
case of the anomalous skin effect.

The real and imaginary parts of the electric field in the vicinity
of the boundary are presented in Figs. 3. The curves depicted in
Fig. 3 correspond to the following parameters: $\varepsilon=10^{-4}$
($\alpha=900$, $\Omega=1000$) for curves
1,$\varepsilon=3\cdot10^{-4}$ ($\alpha=100$, $\Omega=333$) curves 2,
and $\varepsilon=9\cdot10^{-4}$ ($\alpha=11$, $\Omega=111$). This is
an anomalous case. All the curves are considered near plasma
resonance, i.e. the value $\gamma=1$ and $\omega=\omega_p$.

Figure 3a shows the real part of the electric field $\Re e_d(x)$,
which corresponds to the discrete spectrum. An increase in the
anomaly parameters leads to a drastic decrease in the depth of
penetration of the electric field deep into the electron plasma.

Figure 3b depicts the real part of the electric field $\Re e_c(x)$,
which corresponds to the continuous spectrum. In this case, the
electric-field amplitude |$|\Re e_c(x)|<1,2\cdot10^{-9}$ is
considerably smaller that that for the real part due to the discrete
spectrum.

The real part of the electric field $\Re e_c(x)$, which is
associated with the continuous spectrum, is nine orders of magnitude
smaller than the real part of the field $\Re e_d(x)$ corresponding
to the discrete spectrum.

Therefore, the real part of the electric field in the vicinity of
the plasma boundary is actually determined by the discrete spectrum.

The imaginary part of the electric field $\Im e_d(x)$, which is
associated with the discrete spectrum, is shown in Fig. 3c. As the
anomaly parameter increases, the depth of penetration of the
imaginary part of the electric field deep into the plasma decreases
slowly in contrast to the depth of penetration of the real part.

Figure 3d presents the imaginary part of the electric field $\Im
e_c(x)$ which corresponds to the continuous spectrum. Let us compare
the imaginary part of the electric field $\Im e_c(x)$ with the real
part of the field $\Re e_c(x)$. It can be seen that the amplitude
has the same order of magnitude: $|\Im e_c(x)|<1,2\cdot 10^{-9}$.
Therefore, the imaginary part of the electric field at the
aforementioned values of the parameter, in actual fact, is also
determined by the discrete spectrum. However, the imaginary part of
the electric field corresponding to the discrete spectrum is nine
orders of magnitude bigger than the real part of the electric field
corresponding to the discrete spectrum.

As can be seen from the plots presented in Fig. 3, the contribution
of the discrete spectrum at the aforementioned values of the
parameter to the electric field is considerably larger than the
contribution of the continuous spectrum. Thus, the above analysis of
the electric field strength has demonstrated that, in the case of
the anomalous skin effect, the electric field strength is determined
in the vicinity of the boundary by the discrete spectrum.

Figure 4 the modulus of the electric field in the case of the
anomalous skin effect. On the $X$ axis is taken logarithmic scale.
The curves depicted in these figures correspond to the following
parameters: $\varepsilon=3\cdot10^{-3}$ and $\gamma=5$ ($\alpha=5,
\Omega=1666$) for curve 1, $\varepsilon=3\cdot10^{-4}$ and
$\gamma=1$ ($\alpha=100, \Omega=333$) for curve 2. It can be seen
from Fig. 4 that with anomaly parameter increases in 20 times the
modulus of the electric field in the anomalous case decreases one
order of magnitude more rapidly.

{\bf Conclusions}

Thus, in this paper, we have demonstrated that the electric field
and the distribution function of electrons for the skin-effect
problem are determined by their particular solutions. These
solutions are the sums of the solutions corresponding to the
discrete spectrum (decreasing particular solutions) and the
continuous spectrum (solutions of the integral type) and that one
particular solution disappears depending on the anomaly parameter.

It has been established that the zeros of the dispersion functions
are necessary for the analytical solution of the problem and, in
particular, for deriving the electric field and the distribution
function of electrons in the explicit form in the half-space.

The analysis performed in this work has demonstrated that, in the
case of the anomalous skin effect, the electric field in the skin
effect problem is predominantly determined by the discrete spectrum.
The real part of the electric field which corresponds to the
continuous spectrum is eight orders of magnitude smaller than the
real part of the electric field corresponding to the discrete
spectrum.

The imaginary part of the electric field at any anomaly parameters
is considerably bigger than the real part. For $\varepsilon=10^{-4}$
the amplitude of imaginary part is four orders of magnitude bigger
than the amplitude of real part.

\emph{Acknowledgments:} I thank prof. A.V. Latyshev and prof. A.A.
Yushkanov for help with the manuscript.

\begin{figure}[b]
\begin{center}
\includegraphics{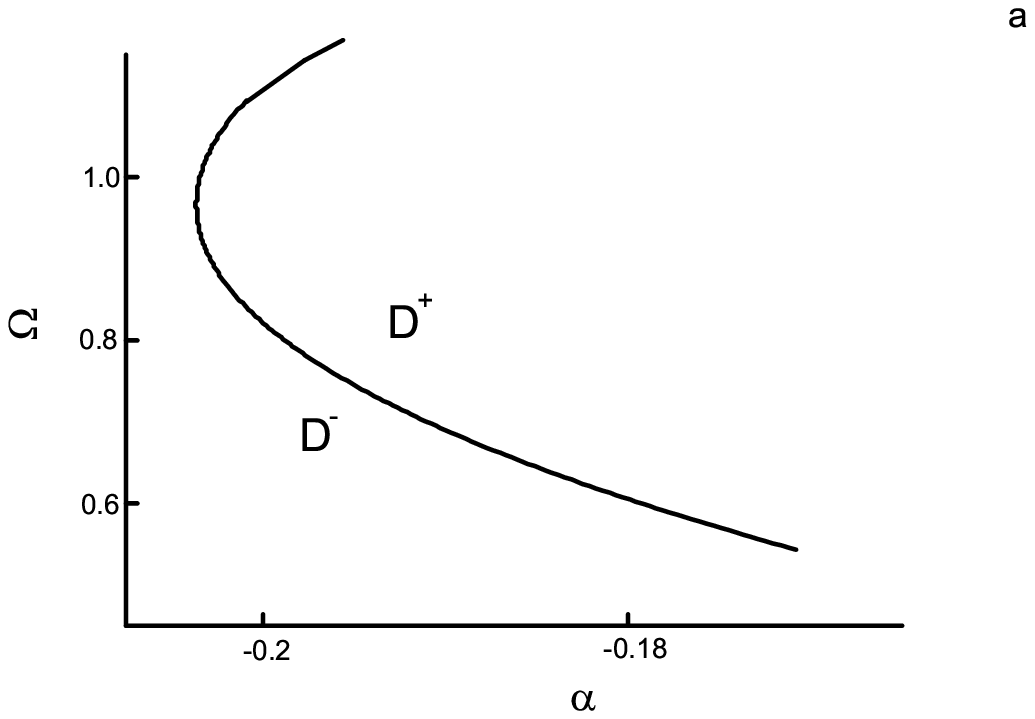}
\end{center}
\begin{center}
{\bf Fig. 1 a. Domains $D^+$ in the $(\alpha,\Omega)$ plane.}
\end{center}
\end{figure}

\begin{figure}[b]
\begin{center}
\includegraphics{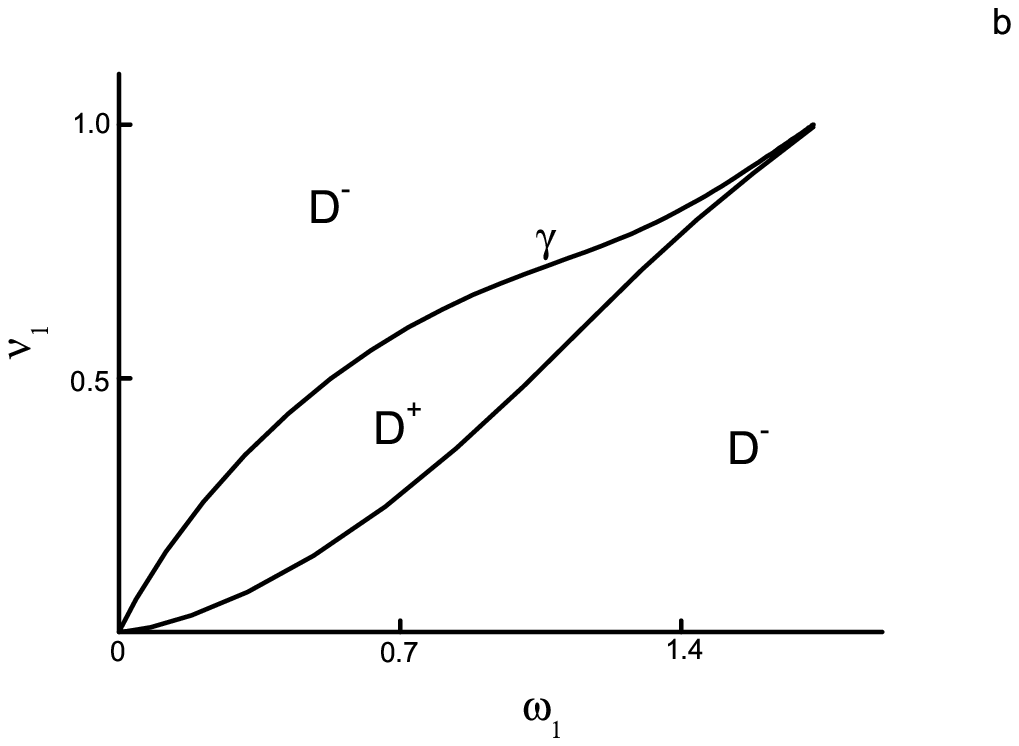}
\end{center}
\begin{center}
{\bf Fig. 1 b. Domains $D^+$ in the $(\omega_1,\nu_1)$ plane.}
\end{center}
\end{figure}

\begin{figure}[b]
\begin{center}
\includegraphics{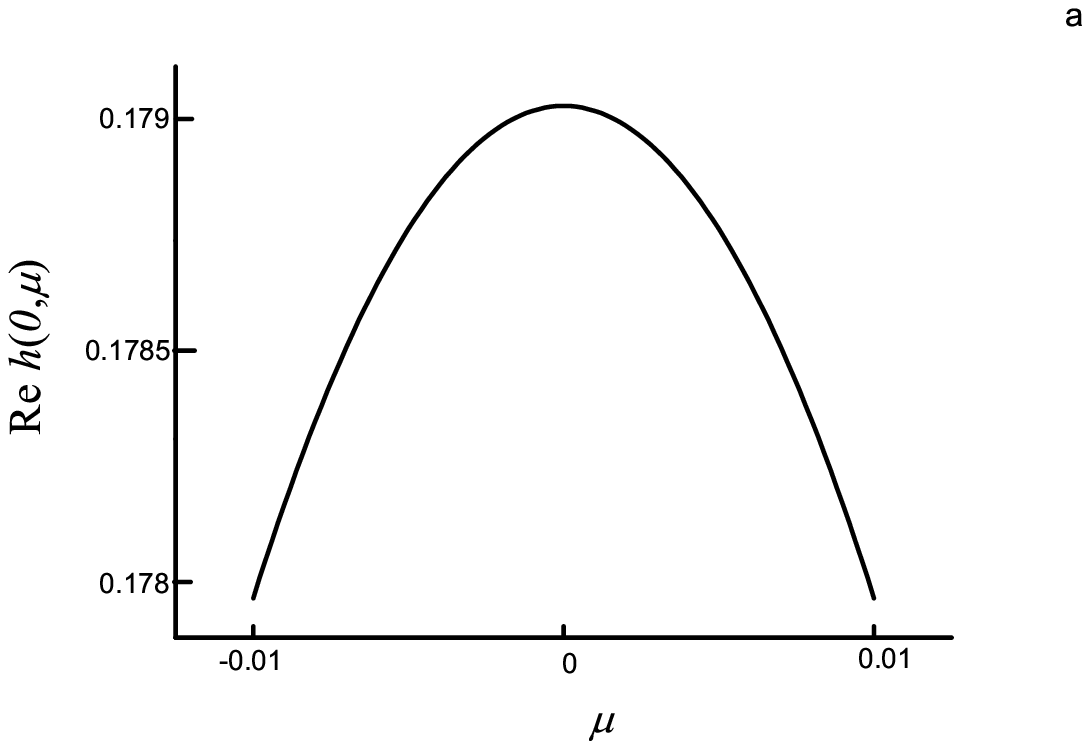}
\\
{\bf Fig. 2a. The real part of the distribution function}
\end{center}
\end{figure}

\begin{figure}[b]
\begin{center}
\includegraphics{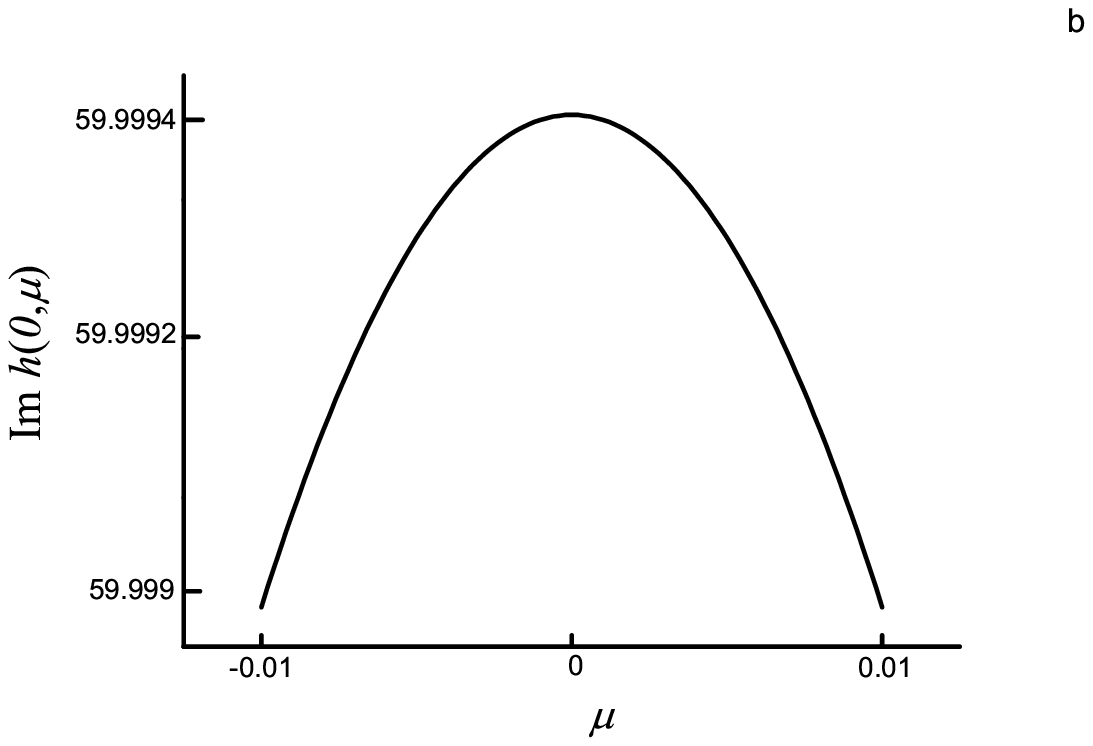}
\\
{\bf Fig. 2b. The imaginary part of the distribution function}
\end{center}
\end{figure}

\begin{figure}[b]
\begin{center}
\includegraphics{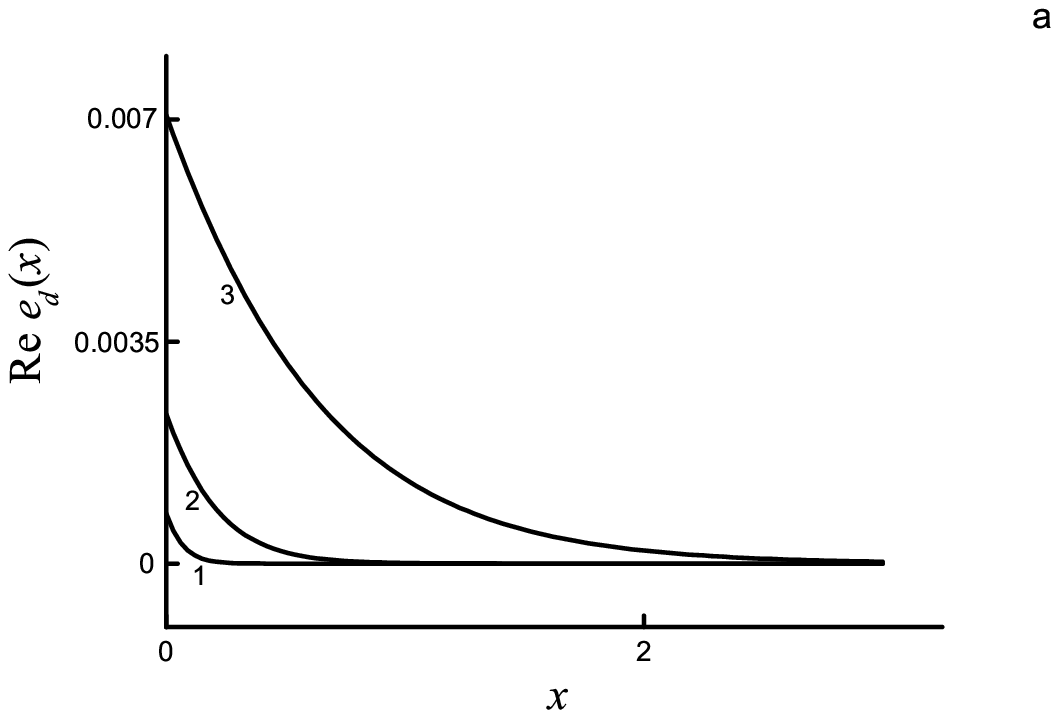}\\
{{\bf Fig. 3a. The real part of the electric field for discrete
spectra.}}
\end{center}
\end{figure}

\begin{figure}[b]
\begin{center}
\includegraphics{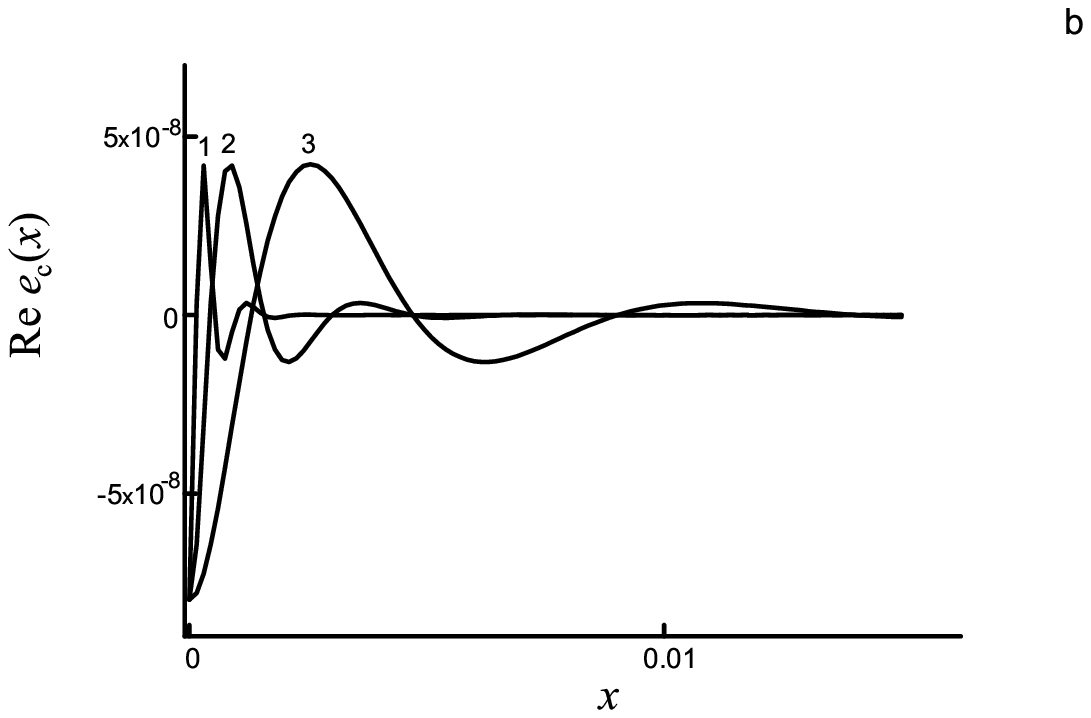}
\\{{\bf Fig. 3b. The real part of
the electric field for continuous spectra.}}
\end{center}
\end{figure}

\begin{figure}[b]
\begin{center}
\includegraphics{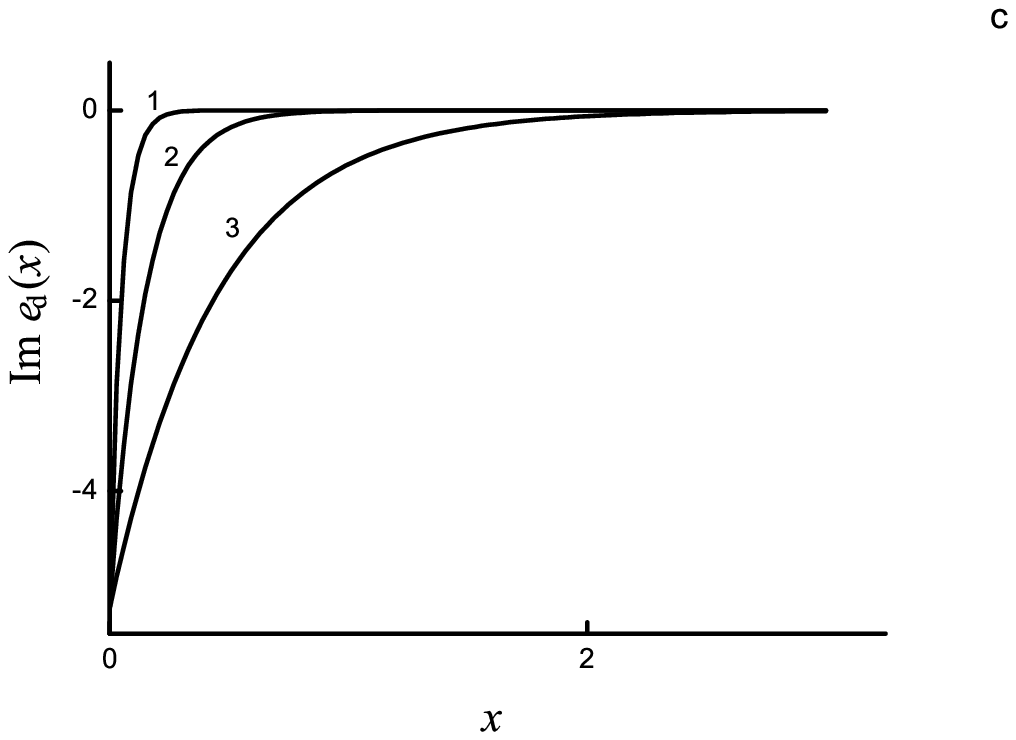}\\
{{\bf Fig. 3c. The imaginary part of the electric field for the
discrete spectra.}}
\end{center}
\end{figure}

\begin{figure}[b]
\begin{center}
\includegraphics{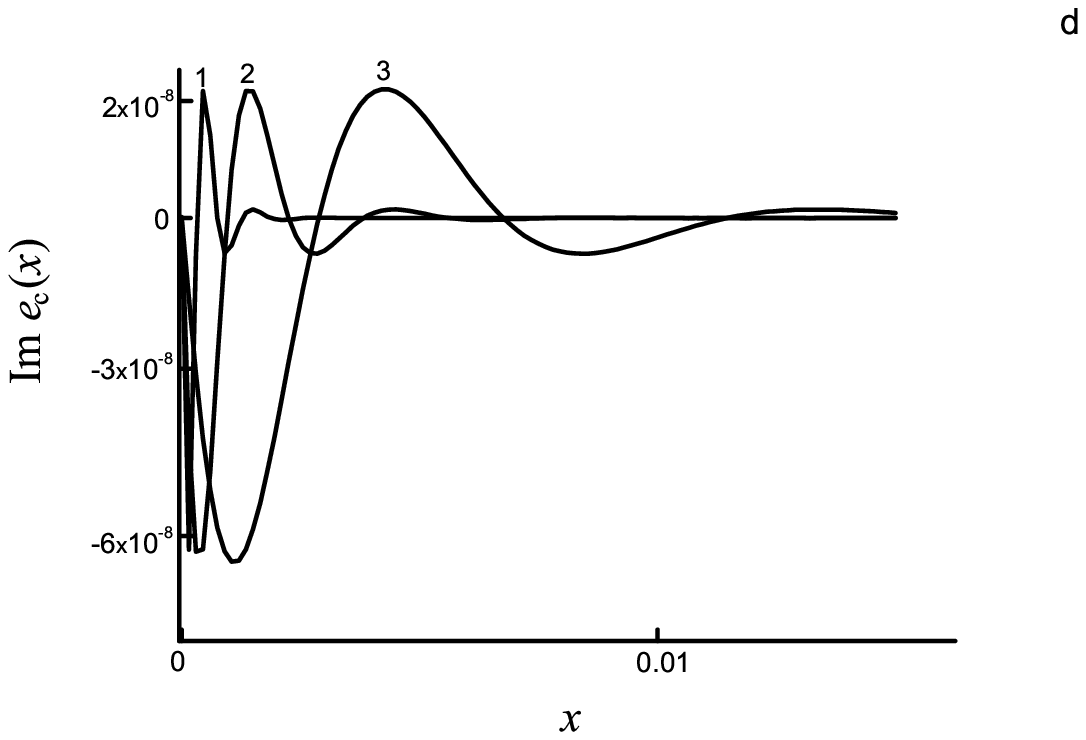}\\
{{\bf Fig. 3d. The imaginary part of the electric field continuous
spectra.}}
\end{center}
\end{figure}

\begin{figure}[b]
\begin{center}
\includegraphics{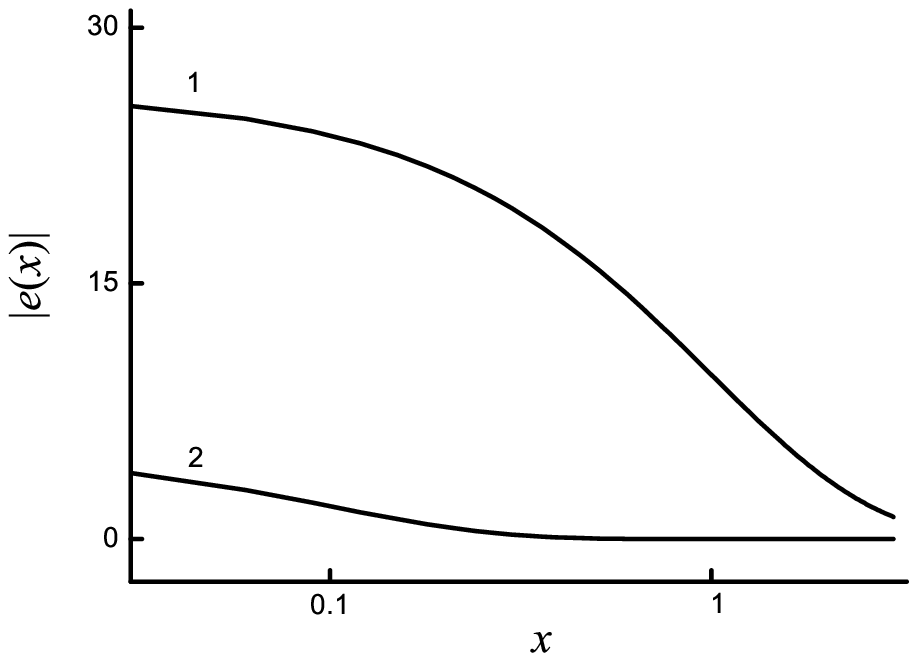}\\
 {\bf Fig 4. The modulus of the electric field}
\end{center}
\end{figure}

\end{document}